\documentclass[aps,pra,twocolumn,superscriptaddress]{revtex4-1}
\usepackage{times}
\usepackage{graphicx}
\usepackage{hyperref}
\usepackage{array}
\usepackage{amsmath}
\usepackage{amssymb}
\usepackage{mathrsfs}
\usepackage{color}
\usepackage{soul,xcolor}
\setstcolor{red}
\usepackage[utf8]{inputenc}
\usepackage[english]{babel}
\usepackage[utf8]{inputenc}
\usepackage[T1]{fontenc}    
\usepackage{amsmath,amsthm,amssymb,amsfonts, mathtools}
\usepackage{graphicx}
\usepackage{braket}

\begin{document}
\title{Decoherence in the three-state quantum walk}

\author{Luísa Toledo Tude}
\affiliation{Instituto de F\'\i sica ``Gleb Wataghin'', Universidade Estadual de Campinas, Campinas, SP, Brazil}
\affiliation{School of Physics, Trinity  College  Dublin,  Ireland}
\author{Marcos C. de Oliveira }
\email{marcos@ifi.unicamp.br}
\affiliation{Instituto de F\'\i sica ``Gleb Wataghin'', Universidade Estadual de Campinas, Campinas, SP, Brazil}

\date{\today}

\begin{abstract}

Quantum walks are dynamic systems with a wide range of applications in quantum computation and quantum simulation of analog systems, therefore it is of common interest to understand what changes from an isolated process to one embedded in an environment. In the present work, we analyze the decoherence in a three-state uni-dimensional quantum walk. The approaches taken into consideration to account for the environment effects are phase and amplitude damping Kraus operators, unitary noise on the coin space, and broken links.  
    
\end{abstract}

\pacs{}
%\keywords{}

\maketitle
    
\section{Introduction}
Quantum walks were initially developed to be the quantum counterpart of the random walks, but soon it was realized that they can be used as a tool to develop faster quantum search algorithms\footnote{For a complete description on quantum walks and search algorithms, see \cite{PortugalBook} and references therein}. This is basically due to the fact that its standard deviation scales as $\sigma \propto t$ which is quadratically faster than the classical case, where $\sigma \propto \sqrt{t}$. Quantum walks can be subdivided in two main groups, discrete and continuous time. Here we focus our attention on the discrete-time quantum walk (DTQW). 
The key element that differentiates the DTQW in the line from the simple random walk is that a quantum version of the coin is taken into account. In other words, this means that while in the random walk a coin is thrown to decide if the next step of a walker will be to the right or to the left, in the quantum case one considers a "coin" --- typically described by the Hadamard gate --- that can be in a superposition of heads and tails, causing the walker to displace in a superposition of right and left steps. The three-state quantum walk is similar to the regular one-dimension walk, but accounting for an additional probability that the walker will remain in the same site during the time step. A typical signature of this extra conditional assignment is the possibility of localization of the walker, contrasting with the standard two-sided coin conditioned evolution  \cite{3QW1,3QW3e4,3QWmatriz}.

With the increase of interest in the field of quantum computation, the theoretical and experimental domain of quantum walks \cite{Gong948} became a key ingredient to the performance of quantum search algorithms. Since quantum computers are physical objects, they are always subjected to some level of noise and dissipation. Therefore, dealing with decoherence is inevitable to build quantum computers that will perform quantum search outperforming classical search algorithms. Decoherence is a key element to understand the limit between classical and quantum phenomena, and there is already an expressive literature on that for two-state quantum walks (See, e.g., \cite{DecoherenceCanBeUsefulinQW}).  A  three-state quantum walk has a richer dynamical structure, which could be explored for quantum simulation of several systems in both condensed matter \cite{AndersonLoc}, and high energy physics \cite{PhysRevA.81.062340}. Since those systems are generally not isolated, an investigation of the effects of decoherence in the three-state quantum walk is in order.   The evolution of a system under decoherence is not necessarily described by unitary operators, therefore one could use external interactions to control a new class of evolutions that lead to different behaviors of the walk.
In this paper, we investigate the effects of decoherence in the three-state quantum walk. Since it can be physically introduced in the system by many different phenomena, depending of the actual physical implementation, we give a general account for such decoherence effects by introducing different mathematical approaches.

This work is subdivided as follows. In  Sec. II we  present a brief overview of the three-state quantum walk in an infinite line. In each of the subsequent three sections, we consider a distinct method of accounting for decoherence. In Sec. \ref{sec:Kraus}, we introduce the Kraus operator that can be used to model phase and amplitude damping decoherence in qutrits. In Sec. \ref{sec:Unise} we investigate decoherence by unitary noise, and at last, in Sec. \ref{sec:BrokenLinks} we analyze decoherence by broken links. Sec. VI is dedicated to the final remarks and conclusions. It is important to point out that those are not the only methods for simulating decoherence, for more information about other methods to implement decoherence in discrete and continuous-time quantum walks we refer to \cite{reviewDecoherence}.

\section{Three-state quantum walk}\label{sec:3QW}
The two-state quantum walk on the line is the quantum version of the simple random walk. Its dynamics can be described by two operators -- one representing the coin toss and the other, the shift of the walker on the line. The difference between the random walk and its quantum version is that in the second case the coin toss does not give an exclusive classical result such as heads or tails, but a superposition of both. In that way, instead of taking a step to the right or to the left, the walker step is a superposition of both directions. 

\begin{figure}[ht!]
    \includegraphics[width=.9\linewidth]{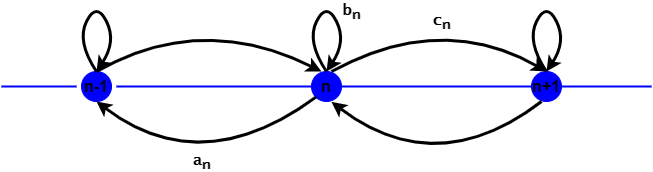}
\caption{Diagram of three-state quantum walk.The coefficients $a_n$, $b_n$ and $c_n$ correspond to the left ($L$), no movement ($S$) and right($R$) chiralities, respectively.}
\label{fig:3QWdiagrama}
\end{figure}

The three-state quantum walk, also known as lazy quantum walk, is analogous to the two-state quantum walk, but with an additional degree of freedom on the "coin" space that accounts for the possibility that the walker does not move in a time step. At each time step the walker flips a "three-sided quantum coin" that falls in a superposition of its three possible states, making the walker's position state, $\ket{n}$, evolve to a superposition of three possible states, $\ket{n-1}$, $\ket{n}$, and $\ket{n+1}$.  Figure \ref{fig:3QWdiagrama} illustrates the possible steps of a walker that occupies the nth site of the lattice.

The system is composed by a coin and a walker, therefore its Hilbert space is written as $\mathcal{H} = \mathcal{H}_C \otimes \mathcal{H}_P$, where $\mathcal{H}_C$ is the "coin" Hilbert space and $\mathcal{H}_P$ the Hilbert Space associated with the positions of the walker in the one-dimensional infinite lattice. The state of the system at anytime can be described as a spinor
\begin{equation}
    \ket{\psi(t)} = \sum^{\infty}_{n = -\infty} 
    \begin{pmatrix}
    a_n (t)\\
    b_n (t)\\
    c_n (t)
    \end{pmatrix} 
    \ket{n},\label{comp}
\end{equation}
where $a_n$, $b_n$, and $c_n$ are the wave components correspondent to the three possibles states of the "coin". Each time step of the Quantum Walk dynamics is composed by two  unitary operations. A rotation in the coin (chirality) space ($C$), followed by a shift (Sh) operation. The operator we will consider here to represent the action of the "coin" is
\begin{equation}
    C =  \frac{1}{3}   \begin{pmatrix}
-1&2&2\\
2&-1&2\\
2&2&-1
\end{pmatrix}, 
\end{equation}
known as Grover coin, and the shift operator is
\begin{equation}
    \begin{aligned}
        \text{Sh} &= \sum^{\infty}_{n = - \infty}\ket{n-1}\bra{n}\otimes \ket{L}\bra{L}\\
        &+\sum^{\infty}_{n = - \infty}\ket{n}\bra{n}\otimes \ket{S}\bra{S}\\
        &+\sum^{\infty}_{n = - \infty}\ket{n+1}\bra{n}\otimes \ket{R}\bra{R}.
    \end{aligned}
\end{equation}
Therefore, using these two unitary operators, the dynamics can be summarized to 
 \begin{equation}
    \ket{\psi(t)} = (\text{Sh} (C \otimes \mathbb{I}))^t \ket{\psi(0)} = U^t\ket{\psi(0)},\label{eq:dynamicsQW}
 \end{equation}
 where $\mathbb{I}$ stands for the identity in position space, and $t$ is the time parametrized as the number of time steps. Using the density matrix notation, equation \ref{eq:dynamicsQW} is equivalent to
 \begin{equation}
     \rho(t) = U^t \rho_0 (U^{\dagger})^t,
 \end{equation}
 where $\rho_0 = \ket{\psi(0)} \bra{\psi(0)}$. The interference between the states will generate a probability distribution of position completely different from the classical. This can be clearly seen as we write the evolution of the global chirality distribution (GCP) \cite{GCP, tude2020temperature},
 \begin{eqnarray}
        \begin{pmatrix}
    P_L (t+1)\\
    P_S (t+1)\\
    P_R (t+1)
    \end{pmatrix} 
    &=&\frac{1}{9}\begin{pmatrix}
    1 &4&4\\
    4&1&4\\
    4&4&1
    \end{pmatrix}
    \begin{pmatrix}
    P_L (t)\\
    P_S (t)\\
    P_R (t)
    \end{pmatrix}- \frac{\mathbb{R}[Q_1(t)]}{9}
    \begin{pmatrix}
    4\\
    4\\
    -8
    \end{pmatrix}\nonumber\\
     &&- \frac{\mathbb{R}[Q_2(t)]}{9}
    \begin{pmatrix}
    4\\
    -8\\
    4
    \end{pmatrix}
    - \frac{\mathbb{R}[Q_3(t)]}{9}
    \begin{pmatrix}
    8\\
    4\\
    4
    \end{pmatrix},\label{eq:evGCP}
\end{eqnarray}
where the GCPs are the total probabilities of having the coin in each state --- right, left, or stay still---, independently of the position,
 \begin{equation}
    \begin{aligned}
    P_L(t) =& \sum^{\infty}_{n = -\infty} |a_n (t)|^2, \\
    P_S(t) =& \sum^{\infty}_{n = -\infty} |b_n (t)|^2,\\
    P_R(t) =& \sum^{\infty}_{n = -\infty} |c_n (t)|^2,\label{eq:GCD3}
    \end{aligned}
 \end{equation}
and the terms $Q_1(t)$, $Q_2(t)$ and $Q_3(t)$, given by
 \begin{equation}
     \begin{aligned}
     Q_1(t) &= \sum^{\infty}_{n = -\infty} a_n(t) b_n^{*}(t);\\
     Q_2(t) &= \sum^{\infty}_{n = -\infty} a_n(t) c_n^{*}(t);\\
     Q_3(t) &= \sum^{\infty}_{n = -\infty} b_n(t) c_n^{*}(t),
     \end{aligned}
 \end{equation}
are responsible for the interference effects of the walk. If the presence of decoherence causes them to completely vanish, equation (\ref{eq:evGCP}) becomes
\begin{equation}
     \begin{pmatrix}
    P_L (t+1)\\
    P_S (t+1)\\
    P_R (t+1)
    \end{pmatrix} 
    =\frac{1}{9}\begin{pmatrix}
    1 &4&4\\
    4&1&4\\
    4&4&1
    \end{pmatrix}
    \begin{pmatrix}
    P_L (t)\\
    P_S (t)\\
    P_R (t)
    \end{pmatrix},
\end{equation}
that describes the evolution of the GCPs as a classical Markovian process. The asymptotic limit of this process gives equal probabilities for each chirality independent of the initial conditions, i.e,
\begin{equation}
    \lim_{n\rightarrow \infty}\begin{pmatrix}
    P_L (n t)\\
    P_S (n t)\\
    P_R (n t)
    \end{pmatrix}
    =\begin{pmatrix}
    1/3\\
   1/3\\
   1/3
    \end{pmatrix}.
\end{equation}

In the next sections we will describe the effects of decoherence in the final probability distribution of positions through different models and analyze how fast they approach the classical limit.
 
\begin{figure}[ht!]
    \centering
    \includegraphics[width=.9\linewidth]{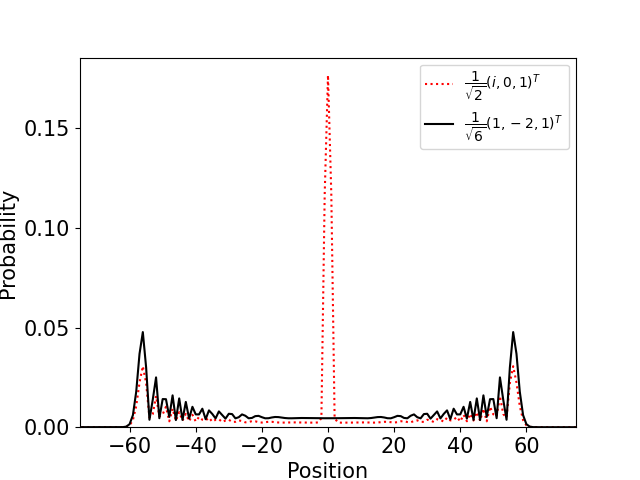}
\caption{Distribution after $100$ time steps. Two different initial conditions were considered one that generates localization, and one that does not, according to Eqs. (\ref{loc}) and (\ref{unloc}).}
\label{fig:3QWpos}
\end{figure}

One of the main differences between the two and three-state quantum walk is that, for some initial conditions, the three state quantum walk can exhibit localization, i.e, a peak on the position distribution around the initial location of the walker. Figure \ref{fig:3QWpos} shows the distribution of displacements of a walker that was initial at the site $0$, with two different initial states of the "coin". The initial conditions were chosen so that we could see the behavior of a walk with and without localization. Those same initial conditions will be used in the rest of the paper, so when we refer to the initial condition that generates localization we will be referring to 
\begin{equation}\label{loc}
    \ket{\psi_0} = \dfrac{1}{\sqrt{2}} \begin{pmatrix}i\\0\\1\end{pmatrix} \ket{0},
\end{equation}
and when we refer to the initial condition that does not exhibit localization, we mean
\begin{equation}\label{unloc}
    \ket{\psi_0} = \dfrac{1}{\sqrt{6}} \begin{pmatrix}1\\-2\\1\end{pmatrix} \ket{0}.
\end{equation}
Note that both states are normalized.

\section{Kraus Operators}\label{sec:Kraus}
The simplest and direct way to introduce decoherence on the walk is by adding extra nonunitary operators, in the form of Kraus operators, $K_j$, to describe the effects of noise and other external effects. Hence, the recurrence relation in respect to the evolution of the system becomes
\begin{equation}
\rho(t+1) = \sum_{j}K_{j} U  \rho(t) U^{\dagger} K_{j}^{\dagger}.\label{eq:KrausEvolution}
\end{equation}
This expression is completely equivalent to considering a unitary evolution on the total space composed by the system and the environment and taking the partial trace of the environment, \cite{Nielsen}.Kraus operators can account for different effects -- here we explore the operators associated with phase and amplitude damping on the coin space, \cite{DoriguelloDiniz2016,Nielsen}. Phase damping is modeled by the Kraus operators of the form $K_{j} = \mathbb{I} \otimes E_j$, where $j = 0,1$, $\mathbb{I}$ is the identity in the position space and the chirality components of the Kraus operators are given by, \cite{KrausOp3QW},
\begin{align}
    E_{0} = \sqrt{1-\gamma} \begin{pmatrix}1&0&0\\0&1&0\\0&0&1
    \end{pmatrix};&&
    E_{1} =\sqrt{\gamma} \begin{pmatrix}1&0&0\\0&\omega&0\\0&0&\omega^2 \end{pmatrix}.\label{eq:Kraus3QWphase}
\end{align}
The parameter $\gamma \in [0,1]$ is the strength of the channel, here called the relaxation parameter, and $\omega = e^{2 \pi i/3}$. This noise process describes the quantum loss of information without loss of energy. Hence, as the system evolves, the information about the phase between the energy eigenstates is lost.

\begin{figure}
  \includegraphics[width=.8\linewidth]{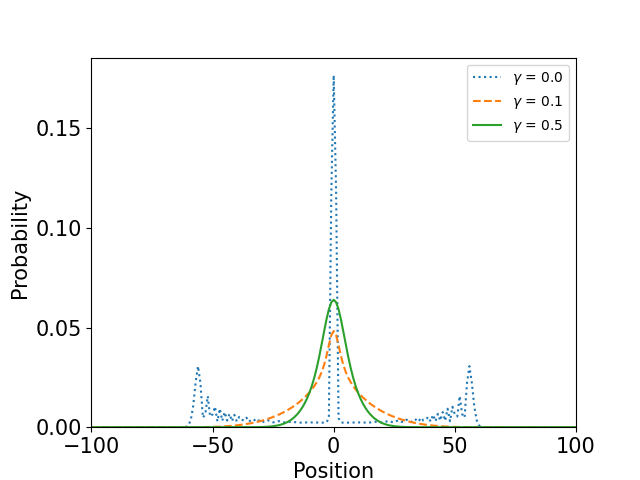}
  \includegraphics[width=.8\linewidth]{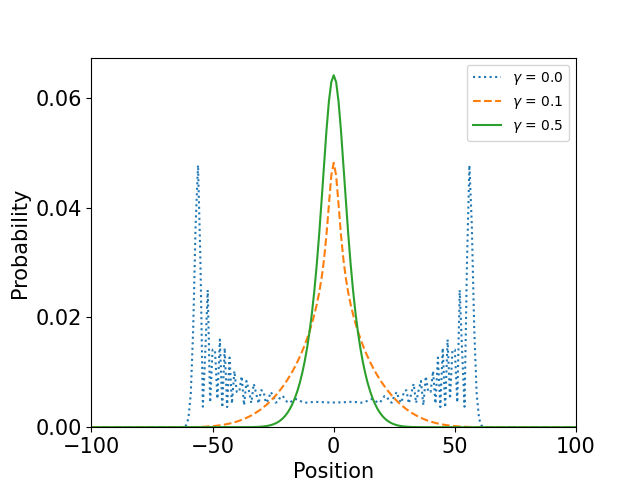}
\caption{Probability distribution of positions of the three-state quantum walk with phase damping after $100$ time steps for different values of the parameter $\gamma$ is displayed for two initial conditions. (a) Top panel for $\ket{\psi(0)} = \dfrac{1}{\sqrt{2}}(i,0,1)^{T}$, and (b) Lower panel for $\ket{\psi(0)} = \dfrac{1}{\sqrt{6}}(1,-2,1)^{T}$. Both situations are equally affected by the dephasing, and the localization is suppressed.}
\label{fig:Kraus3QWPhase}
\end{figure}

The walks with and without localization were simulated considering the phase damping operators, and the result is presented in figure \ref{fig:Kraus3QWPhase}. In both cases the effect of decoherence is similar to the effect in the two-state walk in the sense that there is a transition from quantum to classical (Normal distribution) behavior, \cite{DoriguelloDiniz2016}. The main difference in the case of the three-state walk is that the transition occurs as the relaxation parameter, $\gamma$, increases, but the maximum decoherence is achieved for $\gamma = 0.5$, instead of $1$ and for $\gamma > 0.5$ the distribution starts to transit back to the quantum behavior. Also, note that the dephasing acts equally in the localized and non-localized regimes, and although the normal distribution of the dephased regime, with localized initial condition is around the initial site $0$, the typical localization rapidly disappears. This is expected since any quantum behavior is suppressed equally by the dephasing. Even though localization may occur in any wave-like propagation, here the wave-like behavior is due to the presence of quantum features, in the form of superposition and entanglement of states.

Decoherence is always associated with some kind of informational loss, while the Kraus operators for phase damping introduce loss of phase information, the amplitude damping channel introduces loss of information regarding amplitude, as well as  coherence of the eigenstates. The Kraus operators in this case have a similar format as the ones presented for phase damping decoherence ($K_{j} = \mathbb{I} \otimes E_j$), however in this case, \cite{KrausOp3QW},
\begin{eqnarray}
        E_{0} &=& \begin{pmatrix}1&0&0\\0& \sqrt{1-\gamma}&0\\0&0& \sqrt{1-\gamma}
        \end{pmatrix};\;\; 
        E_{1} = \begin{pmatrix}0&\sqrt{\gamma}&0\\0&0&0\\0&0&0 \end{pmatrix};\nonumber\\
        E_{2} &=& \begin{pmatrix}0&0&\sqrt{\gamma}\\0&0&0\\0&0&0 \end{pmatrix}.\label{eq:Kraus3QWamplitude}
\end{eqnarray}

\begin{figure}[ht]
  \includegraphics[width=.8\linewidth]{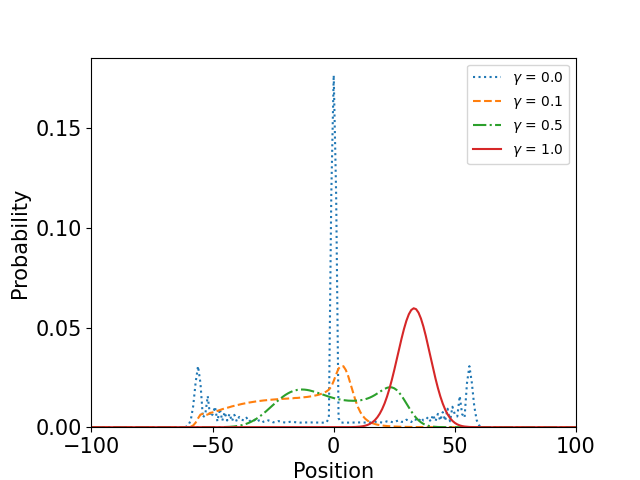}
  \includegraphics[width=.8\linewidth]{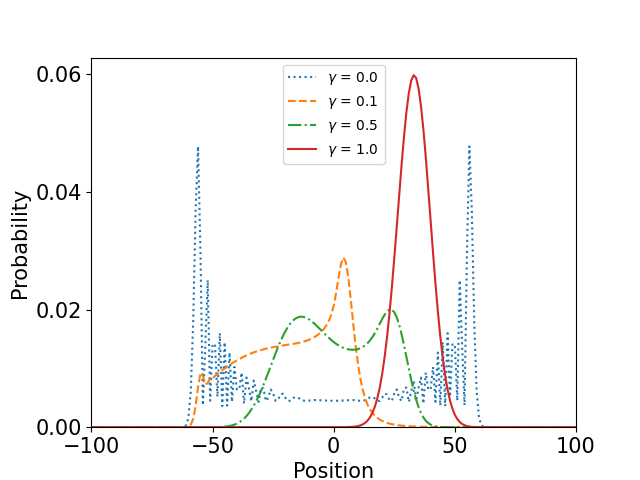}
\caption{Probability distribution of positions of the three-state quantum walk with amplitude damping after $100$ time steps for different values of the parameter $\gamma$ is displayed for two initial conditions. (a) Top panel for $\ket{\psi(0)} = \dfrac{1}{\sqrt{2}}(i,0,1)^{T}$, and (b) lower panel for$\ket{\psi(0)} = \dfrac{1}{\sqrt{6}}(1,-2,1)^{T}$.}
\label{fig:Kraus3QWAmplitude}
\end{figure}

Figure \ref{fig:Kraus3QWAmplitude} shows the resultant distribution of the three-state quantum walk with amplitude damping for the two initial conditions we are considering. The difference between phase and amplitude damping channel effects is in how the transition from quantum to classical behavior occurs. While in the case of phase damping the transitions occur symmetrically, in the amplitude damping  the transition is not symmetric. The walk with strength $\gamma = 0$ represents the walk with no decoherence and as $\gamma$ increases the classical distribution is recovered, but with a shift in the position of the lattice. This is due to the fact that when the decoherence affects the relative phase between the eigenstates, the interference between them is lost,  causing the distribution to behave classically, i.e, respecting the central limit theorem. On the other hand, the amplitude damping channel not only affects the interference, but also the information regarding the mean final position. In any case, despite the asymmetry, the effect is similar to the phase damping, regarding the localization, which is rapidly suppressed with the amplitude damping.

\section{Unitary Noise}\label{sec:Unise}
There is an additional decoherence effect that may occur in quantum walks through unitary processes. Decoherence described by unitary operators can be caused by fluctuations and drifts in parameters (usually associated with phase) of the system Hamiltonian. To consider this type of decoherence in two-state quantum walks, a method was developed in Ref. \cite{UnitaryNoise}. Here we extend the method for three-state quantum walks. It consists in changing the evolution operator $U$ in eq. (\ref{eq:dynamicsQW}) to a unitary operator with a stochastic part. This can be interpreted as a random rotation on the "coin" space at each time step. The dynamics of the system is then given by
 \begin{equation}
     \ket{\psi(t+1)} = \text{Sh}(C e^{i a(t)}\otimes\mathbb{I})\ket{\psi(t)} = Q(t) \ket{\psi(t)}.
 \end{equation}
The operator $a(t)$ is a stochastic and Hermitian operator that acts on the "coin" space. Hence, the new evolution operator, $Q(t)$, is stochastic, but remains unitary. Since the Gell-Mann matrices, ${\bf \lambda} = (\lambda_1, \lambda_2, \lambda_3, \lambda_4, \lambda_5, \lambda_6, \lambda_7, \lambda_8)$, together with the identity form a basis of the chirality space, we can write 
\begin{equation}
    a(t) = \sum_{k = 1}^{8} \alpha_k (t) \lambda_k,
\end{equation}
with $\alpha_k (t)$ being real components of the expansion, that are chosen randomly at each time step.
In this case, the identity does not need to be taken into account because it would only add a global phase to the state. Before simulating the effects of this decoherence on the quantum walk we made the following assumptions on the stochastic operator components $a(t)$, i.e,
\begin{equation}
    \begin{cases}
        \langle \alpha_{k}(t) \alpha_{k^\prime}(t^\prime)\rangle = \delta_{k,k^\prime} \delta_{t,t^\prime}\alpha^2\\
        \langle \alpha_{k}(t) \rangle = 0\ \ \ \ \ \ ;\text{k = x, y, z.}
    \end{cases}
\end{equation}

\begin{figure}[ht!]
      \includegraphics[width=.8\linewidth]{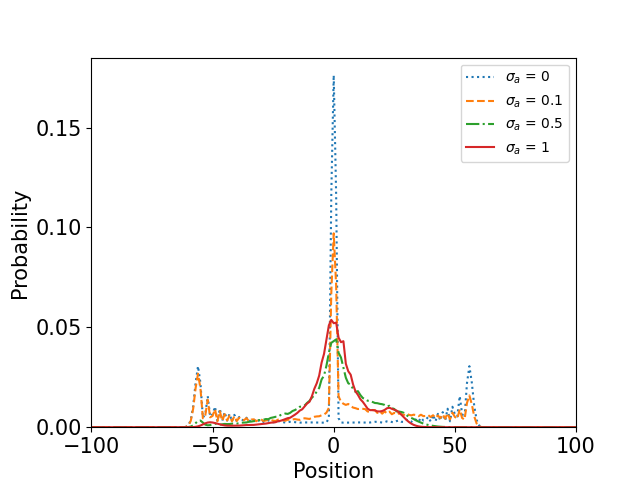}
      \includegraphics[width=.8\linewidth]{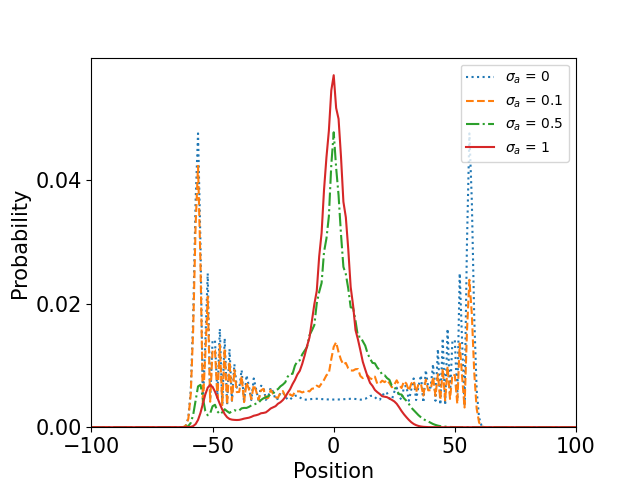}
    \caption{Distribution of position probability of the three-state quantum walk with unitary noise after $100$ time steps for several value of $\sigma_{a}$, for (a) a localized walk with initial state $\ket{\psi(0)} = \dfrac{1}{\sqrt{2}}(i,0,1)^{T}$ in the top panel and (b) a non localized walk, with initial state $\ket{\psi(0)} = \dfrac{1}{\sqrt{6}}(1,-2,1)^{T}$, in the lower panel.}
    \label{fig:Unoise3QW}
\end{figure}

This means that there is no correlation between different components of the operator and between different times and that the probability distributions of $\alpha$ are isotropic. In the simulation, we considered a sampling through a Gaussian distribution with a standard deviation, $\sigma_{a}$, which is varied for each data set. Figure \ref{fig:Unoise3QW} shows the result of the simulations for 4 fixed values of $\sigma_a$. To obtain these distributions, the simulation ran $400$ times and we took the mean of the results. The line that corresponds to $\sigma_{a}=0$ is the limit of the coherent quantum walk, and although the distribution do not seem to reach a Gaussian shape, as $\sigma_{a}$ increases we see a tendency of accumulation on the initial state of the distribution. Therefore, as expected, the stochasticity added to the process diminished the interference effect. We see that, although the random phase affects less drastically the localization for $\sigma_a=0.1$, it affects equally the localized and the non-localized solutions for larger standard deviations of the Gaussian random phase distribution.

\section{Broken Links}\label{sec:BrokenLinks}
When the links between any two sites of the walk have a non-null probability of being broken at each time step, an alternative decoherence source is established \cite{PeriodicMeBrokewnLinks}. If, in a time step, the link is open, the particle cannot move to the neighbor vertices, of the graph where the walk is embedded. In this situation a careful analysis of the recurrence relation for the wave components of the walk, as defined in Eq. (\ref{comp}), must be taken.

The recurrence relation for the regular walk --- i.e, with no broken links --- is
\begin{equation}
 \begin{aligned}
     a_{n}(t+1) &=\frac{1}{3}(-a_{n+1}(t)  +2 b_{n+1}(t) + 2 c_{n+1}(t));
     \\
     b_{n}(t+1) &= \frac{1}{3}(2 a_{n}(t)  - b_{n}(t) + 2 c_{n}(t)); \\
     c_{n}(t+1) &= \frac{1}{3}(2 a_{n-1}(t)  + 2 b_{n-1}(t) - c_{n-1}(t)).
 \end{aligned}
\end{equation}
If the link on the left side of position $n$ is broken, then the upper component of the spinor at $n$ receives a probability flux from $n+1$. To conserve the probability flux, the outgoing flux must be passed to component $c$ at the same site. The resultant expressions are
\begin{equation}
 \begin{aligned}
     a_{n}(t+1) &=\frac{1}{3}(-a_{n+1}(t)  +2 b_{n+1}(t) + 2 c_{n+1}(t));
     \\
     b_{n}(t+1) &= \frac{1}{3}(2 a_{n}(t)  - b_{n}(t) + 2 c_{n}(t)); \\
     c_{n}(t+1) &= \frac{1}{3}(- a_{n}(t)  + 2 b_{n}(t) + 2 c_{n}(t)).
 \end{aligned}
\end{equation}
Analogously the recurrences relations in the case that there is a broken link to the right of the site $n$ are
\begin{equation}
 \begin{aligned}
     a_{n}(t+1) &=\frac{1}{3}(2 a_{n}(t)  +2 b_{n}(t) - c_{n}(t));
     \\
     b_{n}(t+1) &= \frac{1}{3}(2 a_{n}(t)  - b_{n}(t) + 2 c_{n}(t)); \\
     c_{n}(t+1) &= \frac{1}{3}(2 a_{n-1}(t)  + 2 b_{n-1}(t) - c_{n-1}(t)).
 \end{aligned}
\end{equation}
Finally, if both links that connect site $n$ with its neighbors are broken, the relations become
\begin{equation}
 \begin{aligned}
     a_{n}(t+1) &=\frac{1}{3}(2 a_{n}(t)  +2 b_{n}(t) - c_{n}(t));
     \\
     b_{n}(t+1) &= \frac{1}{3}(2 a_{n}(t)  - b_{n}(t) + 2 c_{n}(t)); \\
     c_{n}(t+1) &= \frac{1}{3}(- a_{n}(t)  + 2 b_{n}(t) + 2 c_{n}(t)).
 \end{aligned}
\end{equation}

\begin{figure}[ht!]
    \includegraphics[width=.9\linewidth]{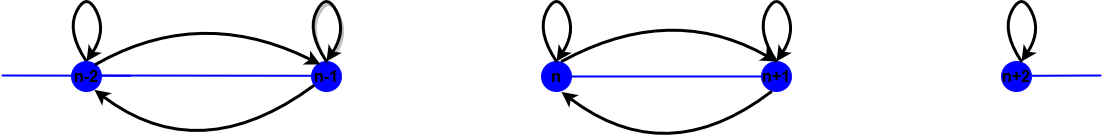}
\caption{Diagram of three-state quantum walk with broken links. At every time step of the process a new set o broken links is randomly generated.}
\label{fig:3QWBrokendiagrama}
\end{figure}

At each time step some links are randomly chosen (with probability p) to be broken. Then, the process evolves following a different recurrence relation to each position, depending if its neighbor links are broken or not. Figure \ref{fig:3QWBrokendiagrama} illustrates the possible steps of a walker on a lattice with some broken links.

The evolution proceeds through unitary operations, however, the operators change randomly according to the topology of the graph. Note that the decoherence comes from a stochastic process --- i.e, the changes on the links of the graph --- that changes the evolution operator, $U$, to another unitary operator, just like in the case of the unitary noise model. The main difference between both cases is that here the noise directly affects the walker space, while in section \ref{sec:Unise} it accounted for decoherences in the coin space.

\begin{figure}
  \includegraphics[width=.8\linewidth]{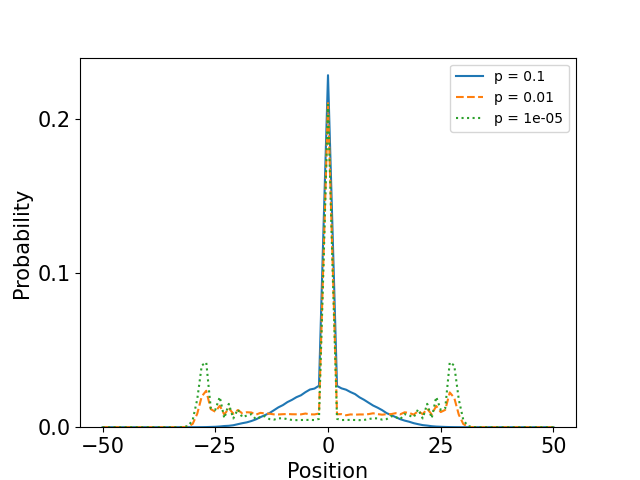}
  \includegraphics[width=.8\linewidth]{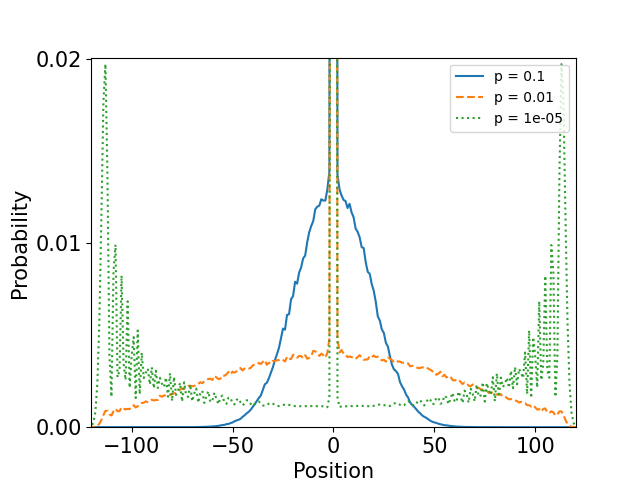}
\caption{Probability distribution of the position of the walker after $50$ time stepes (top panel) and $200$ time steps (lower panel) obtained by the mean values of $1000$ simulations of the three-state quantum walk with broken links for three different probabilities of broken links. The initial condition considered is $\dfrac{1}{\sqrt{2}}(i,0,1)^{T}$.}
\label{fig:BrokenLinks3QWLoc}
\end{figure}

\begin{figure}
  \includegraphics[width=.8\linewidth]{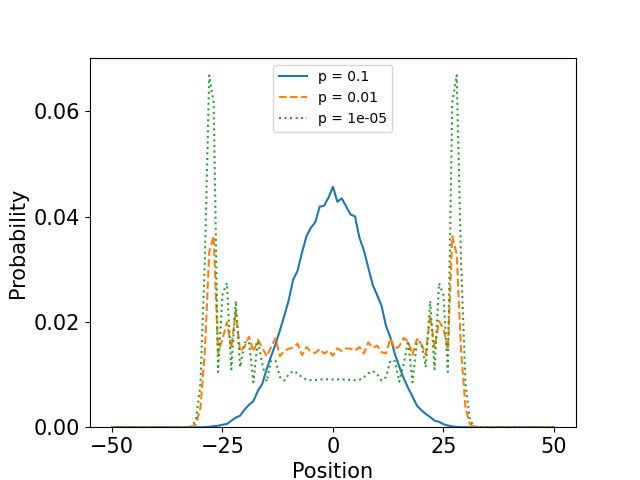}
  \includegraphics[width=.8\linewidth]{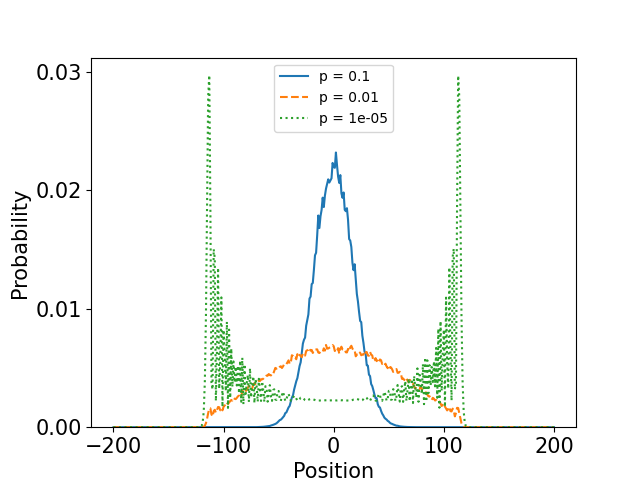}
\caption{Probability distribution of the position of the walker after $50$ time steps (top panel) and $200$ time steps (lower panel) obtained by the mean values of $1000$ simulations of the three-state quantum walk with broken links for three different probabilities of having a broken links. The initial condition considered is $\dfrac{1}{\sqrt{6}}(1,-2,1)^{T}$.}
\label{fig:BrokenLinks3QWNLoc}
\end{figure}

Using the recurrence relations derived above, we simulate the three-state quantum walk with broken links after $50$ and $200$ time steps and for different probabilities of broken links. Figures \ref{fig:BrokenLinks3QWLoc} and \ref{fig:BrokenLinks3QWNLoc} show the mean result of $1000$ simulation runs for the two initial conditions.
%..........
As in the processes analyzed in the previous sections, we see a transition from the quantum to a Gaussian-like  distribution, however in this case and interesting feature differentiates the effect of the decoherence. When the initial condition generates localization, the broken links preserve it, changing only the other regions of the distribution. In figure \ref{fig:BrokenLinks3QWLoc} we can see clearly that outside the localization region the blue and orange curves approach a Gaussian shape and in the central region the three curves present the localized shape. This is due to the fact that the disorder introduced by the random choice of broken links, in fact contributes for localization. However it is not strong enough to imprint a localization profile for arbitrary initial conditions, only for the favorable localized initial condition is that the localization is reinforced.

\begin{figure}[htp]
  \includegraphics[width=.8\linewidth]{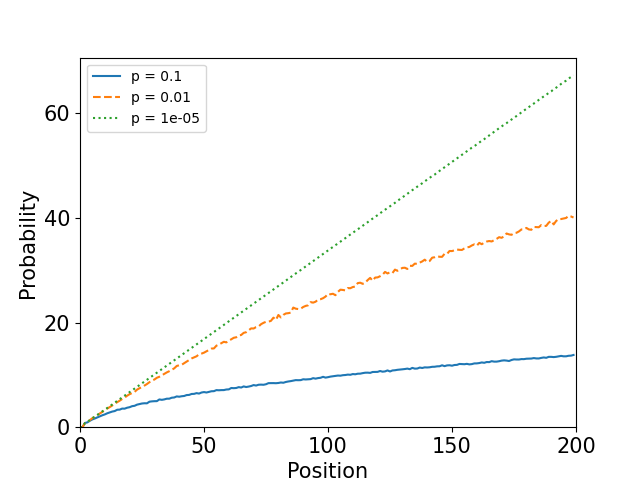}
  \includegraphics[width=.8\linewidth]{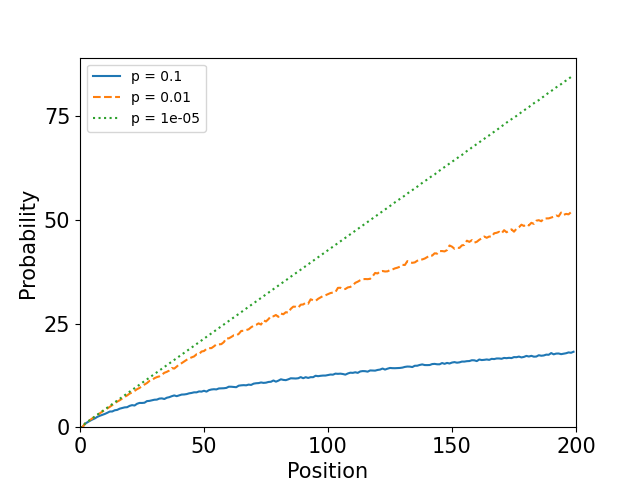}
\caption{Standard deviation of displacement for the three-state quantum walk with broken links for the initial conditions (a) $\ket{\psi(0)} = \dfrac{1}{\sqrt{2}}(i,0,1)^{T}$ in the top panel, and (b) $\ket{\psi(0)} = \dfrac{1}{\sqrt{6}}(1,-2,1)^{T}$ in the lower panel.}
\label{fig:sigma}
\end{figure}

 There is a characteristic time, $t_c$, associated with the transition between quantum to classical behavior that depends on the probability of broken links $p$, \cite{PeriodicMeBrokewnLinks}. At the initial time, the walker is at position $0$, therefore there are only two relevant links to the walk -- the ones connecting position $0$ with $\pm 1$. As the time evolves the wave function spreads through the line covering a range of $\alpha t$ . Hence, the mean number of broken links per time step is proportional to the time, $p \alpha t$. The classical behavior starts to emerge when the mean number of broken links per time step is of order one, so $t_c = \dfrac{1}{p \alpha}$ and for $t>>t_c$ the distribution tends to a Gaussian. The transition is also reflected on the standard deviation, the spread for early times is ballistic, and for $t>>t_c$ the classical spread is dominant. 
This can be observed in figure \ref{fig:sigma}. All three curves start looking like a straight line but as time grows, the ballistic feature stops to be the dominant behavior. This happens first to the walk with a higher probability of broken link. The green curve remains with quantum behavior through all time accounted.

\section{Conclusions}\label{sec:C}

Despite being considered as the quantum counterpart of random walks, quantum walks are not stochastic processes in the classical sense. That is, randomness plays a clear role at each time step of the random walk, in the sense that the result of the coin toss is not predictable and the system's dynamics is irreversible. On the other hand, in the quantum walk, the position of the walker is unknown, but the state of the system is always known. The result of the coin toss is perfectly predictable and the dynamics of the system is governed by a unitary evolution, which means that if the initial state is pure it will remain pure. The randomness of the quantum walk is uniquely due to the measurement process.

However, when the quantum walk is not completely isolated, the picture changes and the lack of knowledge about the environment that contains the system can add randomness to the system. Those decoherence effects are sometimes inevitable with the available technologies nowadays, which makes extremely important to understand what different types of interactions between system and environment have in the process we want to control. In that spirit, this paper is devoted to the analyses of the decoherence effect in the three-state quantum walk.
We analyzed the behavior of the walk for different degrees of relaxation in the case of phase and amplitude damping and observed a Gaussian behavior emerging in both cases, but with the difference that the amplitude damping also generates a lost symmetry in the probability of position. In sections \ref{sec:Unise} and \ref{sec:BrokenLinks} we analyzed a decoherence that causes a random factor in the Hamiltonian. In the first case this randomness is related to the chirality space while in the second it affects the position space. In the case of decoherence by broken links we also notice that the decoherence preserves the localization of the distribution.

Besides its experimental applications, this work also presents a simple way of understanding the quantum to classical limits. As expected, our results suggest that the decoherence attenuates the quantum interference effects that are responsible for the wavy shape of the position distributions and for the linear grow of the the standard deviation of the distribution in time. 
\begin{acknowledgments}
 This work was partially supported by CNPq (Brazil).
\end{acknowledgments}
\bibliographystyle{ieeetr}

\bibliography{Bibliografia}   
\end{document}